# Beyond Silence, Redefining Libraries into Community Hubs: A Comparative Study on Gen Y and Gen Z

Jagan B L
Research Scholar
Dept. of Library and Information Science,
University of Kerala, Thiruvananthapuram
https://orcid.org/0009-0007-2295-797X
jaganbl14061999@gmail.com

Dr. Sudhi S. Vijayan
Assistant Professor & Head
Dept. of Library and Information Science,
University of Kerala, Thiruvananthapuram
https://orcid.org/0000-0002-2228-9031
sudhivijayan@gmail.com

Dr. V. R. Renjith
Assistant Librarian Gr. II
Kerala University Library, University of Kerala,
Thiruvananthapuram, Kerala
https://orcid.org/0000-0002-2389-7338
renjithlib@keralauniversity.ac.in

## Abstract

*The library environment is changing rapidly, and keeping up with the modern era is the only way to provide effective and efficient service to users. Silent libraries and strict librarians have become obsolete; collaborative spaces and community spaces are the future of libraries. The present study aims to identify the willingness and skill level of library professionals to implement new updates in the services, availability of infrastructure to provide community-integrating services as well as to compare the opinions of Gen Y and Gen Z regarding the transformation of libraries. The study used a structured questionnaire to collect data from MLISc postgraduates of the Department of Library and Information Science, University of Kerala, who pursued their studies during the time period 2015-2024. From the population of 200, sample size of 127 is calculated using Krejcie and Morgan table. Collected data is further analyzed with Microsoft Excel and SPSS. Hypotheses were framed and tested with One Sample t-test and Independent Samples t-test. Findings of the study revealed that there is a positive attitude towards the transformation of libraries into community hubs and both Gen Y and Gen Z have the same opinion regarding the change.*

**Keywords:** ***Community Hubs, Community Centers, Collaborative Spaces, Gen-Y, Gen-Z, Millennials.***

## 1. Introduction

The traditional image of quiet libraries with stacks of books and strict librarians are becoming obsolete. Modern era demands libraries to be interactive places which support community engagement programs. Libraries are performing its roles as catalysts for civic activities and participation in the digital age. These changes are challenging the traditional style of libraries as well as library professionals. The sustainability of public library depends on the capability of libraries in performing community services in which diverse user population can gather and conduct society building practices, becoming the center of democracy will increase the value of libraries.

Transformation of services will enhance the public libraries also it will strengthen the democracy and sustainability of the library (Hapel, 2020), community engagement and utilization of library space can harvest support and trust from the public (Ramya & Abdul Majeed, 2025). Libraries always have been a central point of social gathering and played a prominent role in the cultural, socio-political development (Kunhambu & Mudhol, 2012), beyond just book lending and silent rooms libraries should be place were cultural and community programs are conducted (Abu et. al, 2011).

## 2. Previous studies

Technological innovations are transforming the higher education sector, where library act as physical spaces where people connect with knowledge. The transformation of library was studied by **Garoufali & Garoufallou (2022)**. The study aims to understand the opinions of Greek academic librarians regarding the shift of the purpose of libraries. Data for the study is collected using online survey, structured questionnaires are distributed to the 37 academic institutions, and further analysis of the study is done using ANOVA and t-test. Result of the study reveals that majority of the libraries welcomes the transition of the libraries into community places. Most of the libraries already implemented digital hubs but full transformation of libraries are prevented by the structure of the library building.

The participation of the users in the community engagement programme offered by the public libraries in Kerala was studied by **Ramya & Abdul Majeed (2025)**. Libraries have been providing outreach programs to the public, prominent objective of the study is to find the utilization of public library's community engagement services by the different user groups, further the study intend to understand the challenges faced by the users as well as library professionals in using such kind of outreach services. Methodology of the study consists of a well-structured questionnaire which is distributed among the different reader group communities namely Balavedhi, Yuvajanavedhi, Vanithavedhi and Vayojanavedhi in public libraries functioning under the Kerala State Library Council. Chi-square test is used to analyze the collected data and results revealed that most of the users utilize the community services provided by the public libraries also the users suggest that additional facilities such as Wi-Fi, ICT and more e-governance services should be incorporated to the community hubs.

## 3. Methodology

The study's methodology follows an analytical method. The primary information is collected through a structured questionnaire. The previous 10 batches (2015-2024) of MLISc students from the Department of Library and Information Science, University of Kerala, were taken as the population for the study, intending to accommodate Gen Y and Gen Z. From the total population of 200, 126 participants were taken as the sample for the study using the Krecji and Morgan Table (Krejcie & Morgan, 1970). Collected data are further analysed using Microsoft Excel and SPSS software.

## 4. Objectives

- To evaluate whether the traditional quiet library model has become obsolete.
- To understand the preparedness of library professionals in welcoming community library space as a service
- To assess the need of library space planning in accommodating community space services
- To evaluate the need for traditional and modern library functions
- To understand the importance of revision in LIS education to address modern library services
- To recognize the challenges faced while implementing collaborative spaces
- To compare the opinions of Gen Y and Gen Z

## 5. Hypotheses

The following hypotheses were formulated for the present study.

**H1:** Opinions regarding collaborative spaces are equal to the average level

**H2:** There is no significant difference between the opinions of Gen Y and Gen Z.

## 6. Data Analysis

The total population has been categorized based on the different demographic characteristics

### 6.1 Gender-wise analysis of respondents

Respondents are divided into two categories: male and female. The distribution of the respective categories has been given below in Figure 1.

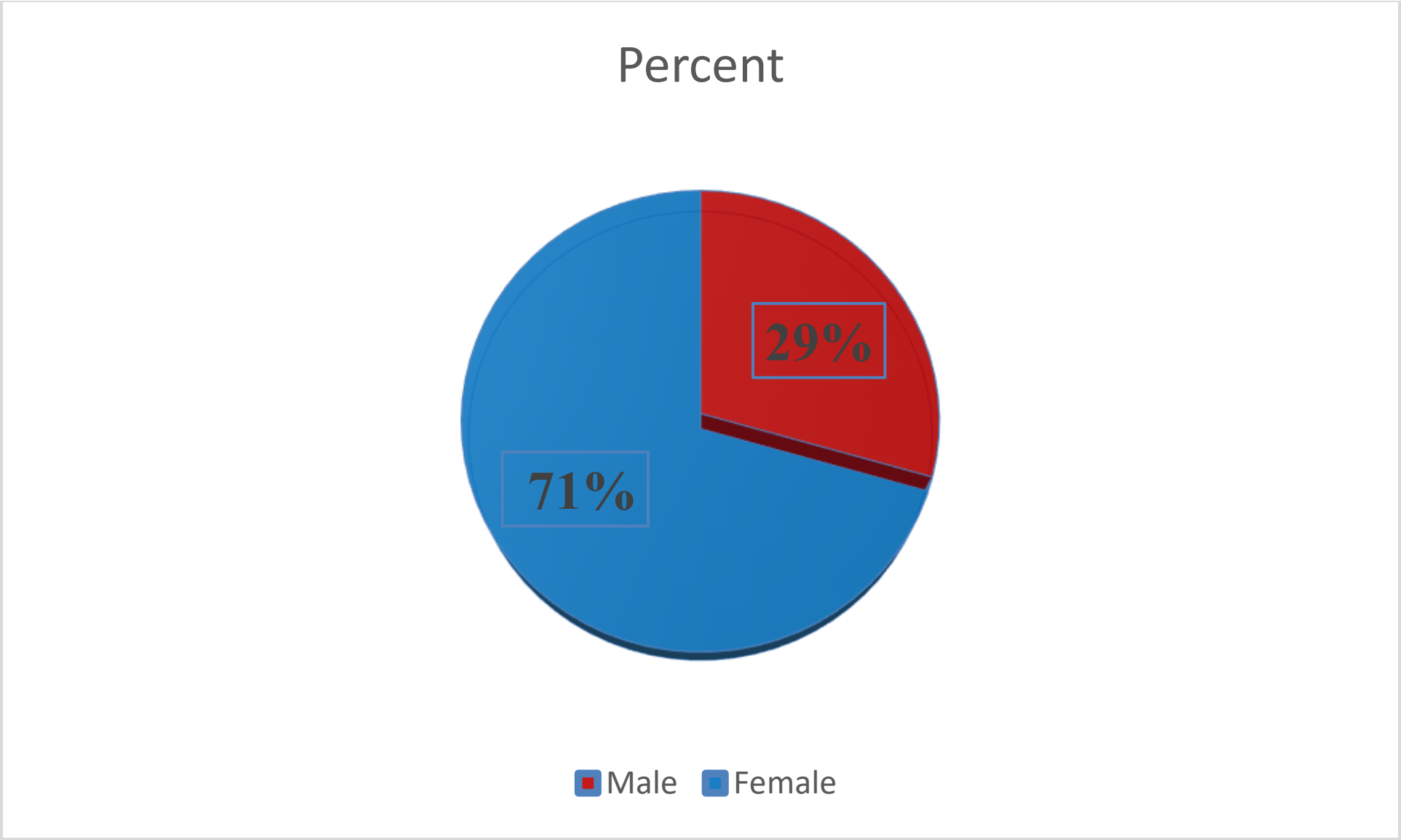

**Figure 1 : Gender**

This figure shows the gender-wise distribution of respondents, revealing a significant gender imbalance with 89 females (70.6%) and 37 males (29.4%). Considering academic Librarians, women tend to dominate the field (Monroe-Gulick, 2024). The figure also indicates female dominance among the respondents.

### 6.2 Genaration-Wise Distribution of Responsents

The population consists of Generation Y (Gen Y) and Generation Z (Gen Z), the year of birth is used to divide the population into Gen Y and Gen Z that is before 1996 and after 1996. dividing respondents into Gen Y and Gen Z. Gen Y is also known as Millennials; the frequency distribution of generations is given below in Figure 2.

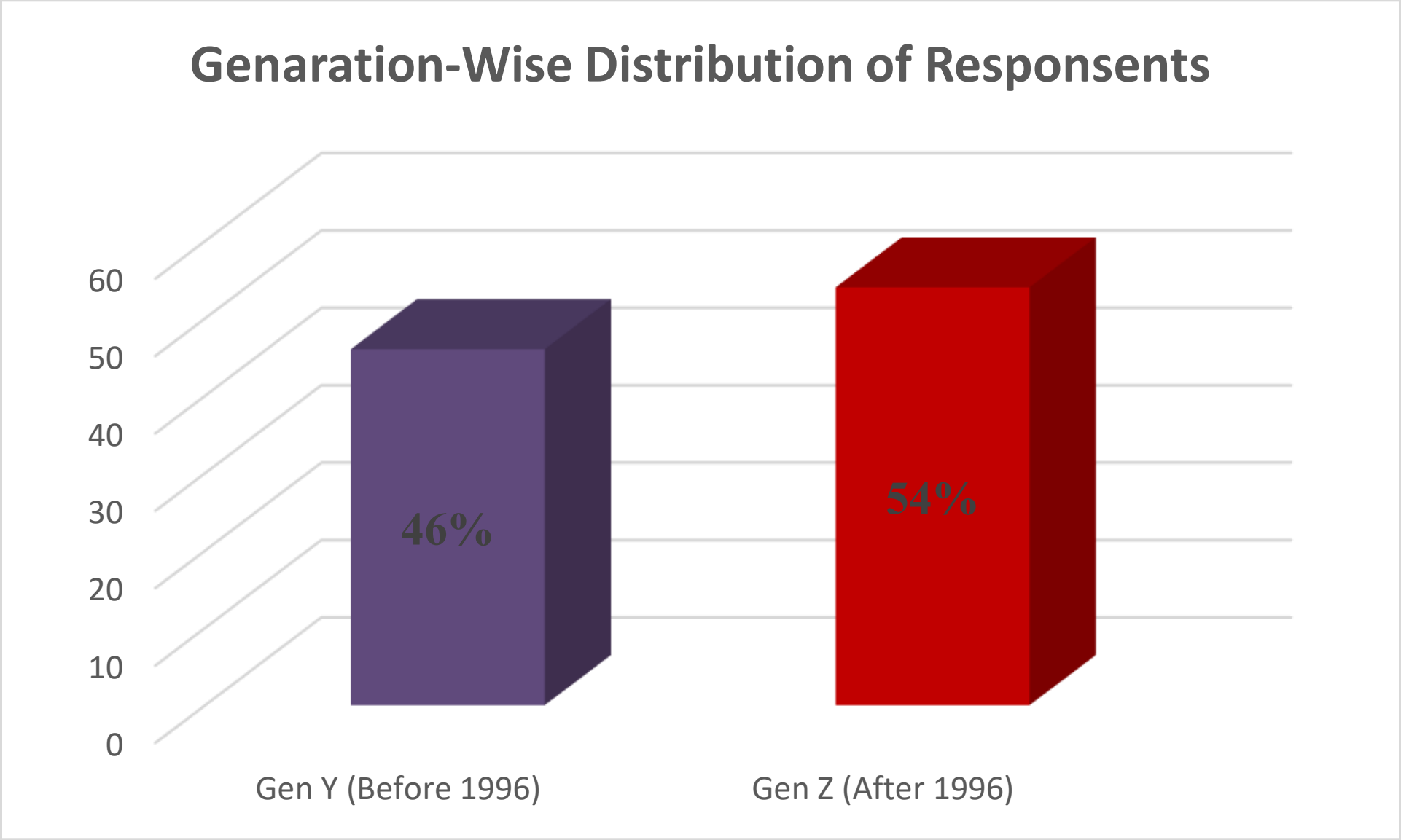

**Figure 2: Genaration-Wise Distribution of Responsents**

This figure indicates the distribution of a 126-person sample that 58 people (46%) classified as" Before 1996" and 68 people (54%) classified as" After 1996." The table structure follows standard frequency analysis format, where frequency represents raw counts, percent and valid percent are identical, indicating there are no missing data and cumulative percent shows the running total reaching 100.

### 6.3 Current Employer Status

The employer details of the respondents are taken using three option which are Academic library, Public library and special library. The frequency of their distribution is given below in Fiure 3.

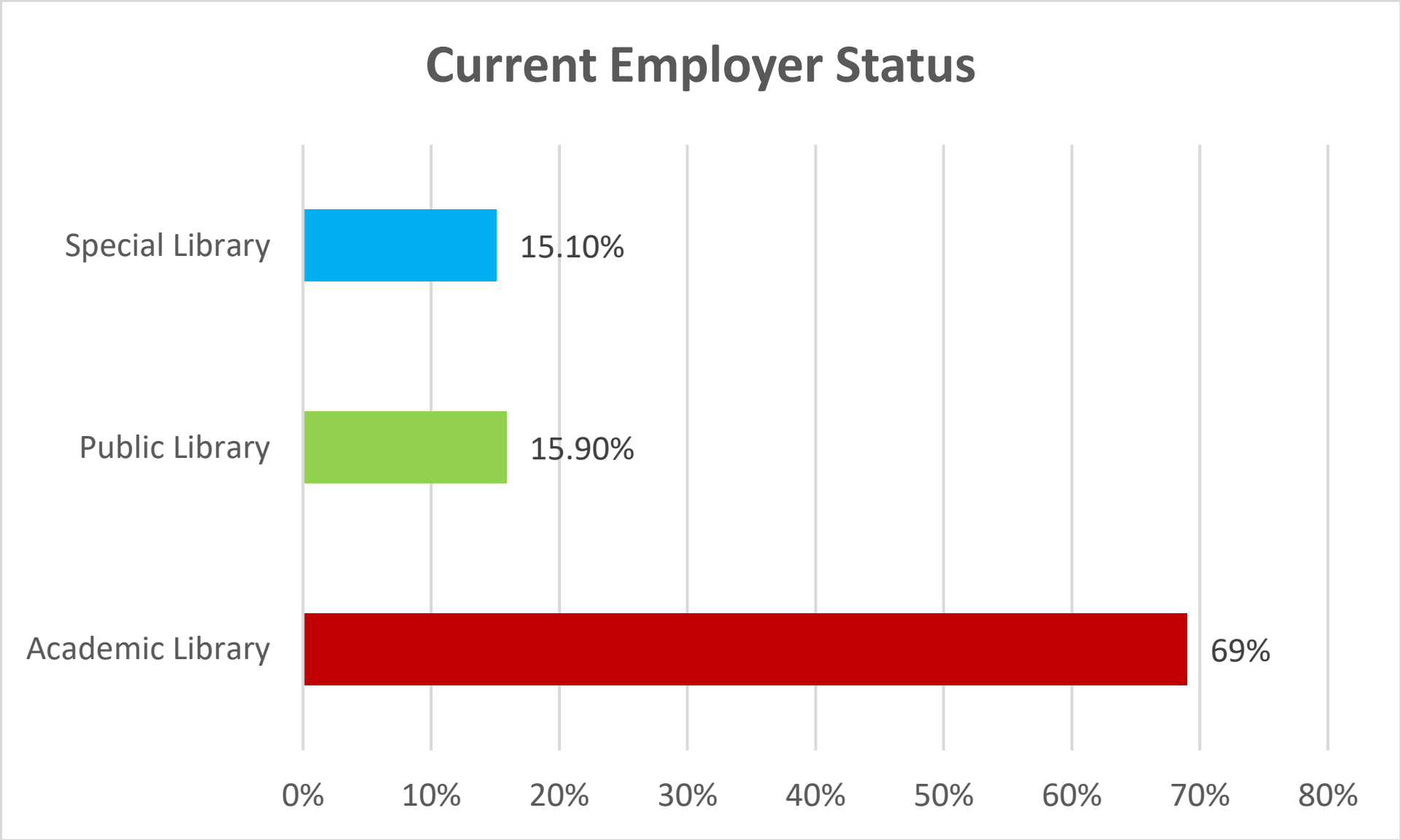

**Figure 3: Current Employment Status**

This frequency distribution table shows the breakdown the type of library where the 126 participants are employed, revealing a strong attention in academic libraries with 87 responses (69%), while public libraries and special libraries have much lower and nearly equal representation with 20 responses ( 15.9%) and 19 responses ( 15.1%) respectively.

### 6.4 Working Experience in the LIS Field

The duration of the respondent's working experience has been measured using a five-point scale, and the frequency of distribution is given below in Figure 4.

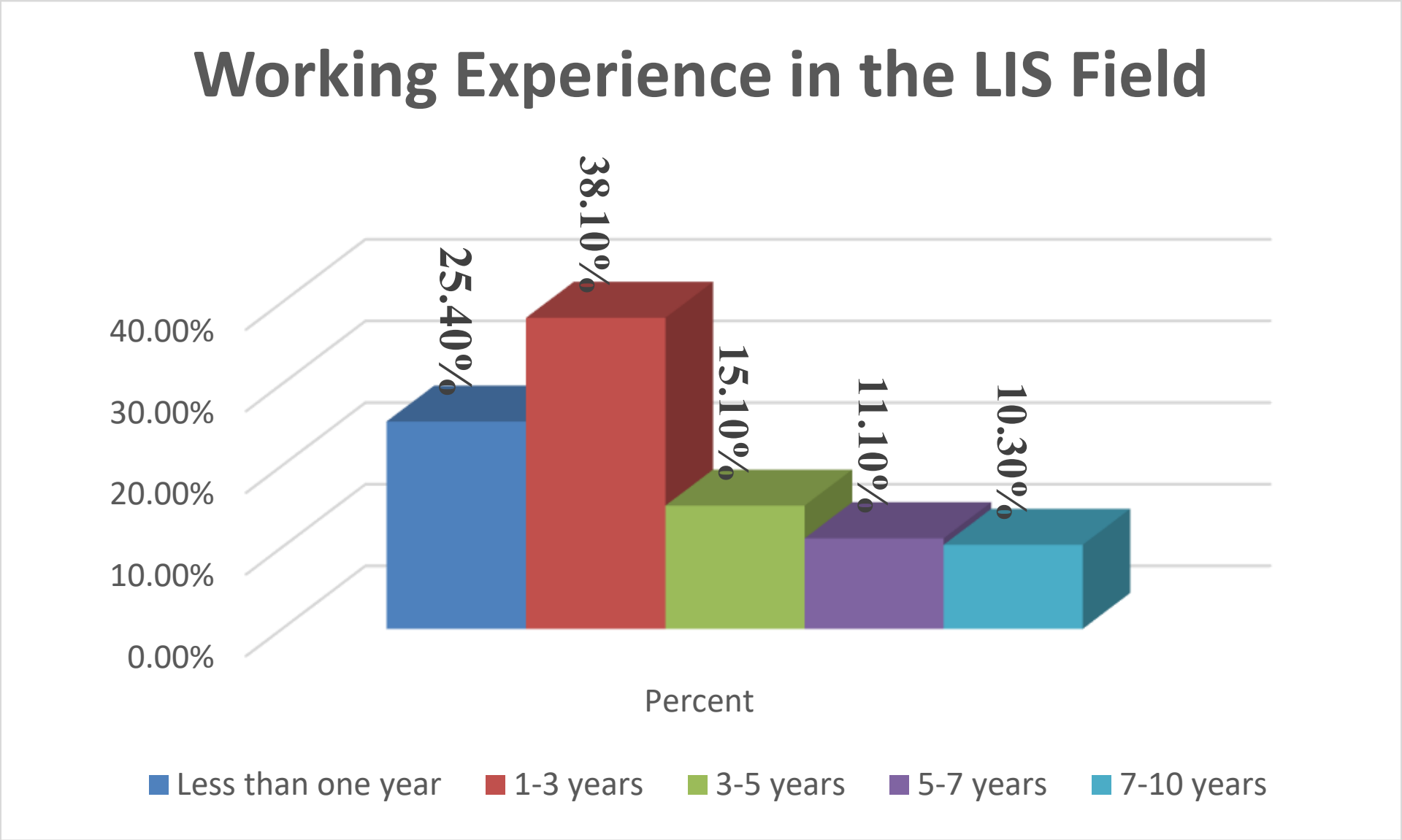

**Figure 4: Working Experience in the LIS Field**

This frequency distribution figure shows the years of experience of 126 samples across five orders, revealing that the majority of the respondents are fairly new to the field of Library service, with 80 responses (63.5%). The largest single group consists of those with 1- 3 times of experience at 48 responses (38.1%), followed by those with lower than one time at 32 responses (25.4%), while the remaining orders show declining frequentness 19 people (15.1%) with 3- 5 times, 14 people (11.1%) with 5- 7 times, and 13 people (10.3%) with 7- 10 times of experience.

### 6.5 Preparedness for the Transformation of Libraries

The preparedness of the respondents regarding the transformation of libraries into community hubs have been measured using five point scale ranging from Very unprepared to Very prepared. The responses are given below in the Figure 5.

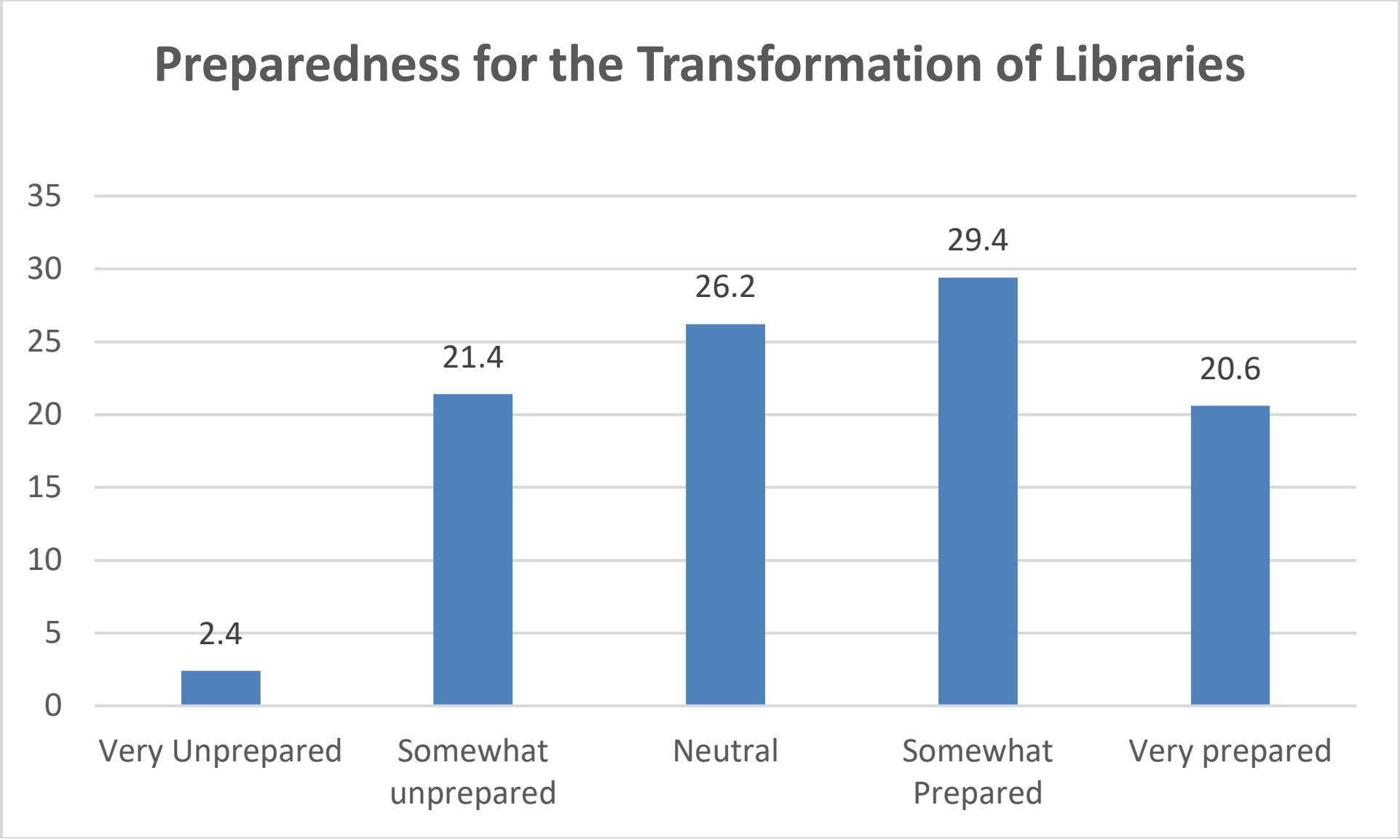

**Figure 5: Preparedness for the Transformation of Libraries**

This survey of 126 respondents reveals mixed opinions about new MLISc graduates' preparedness to manage library transformation into communal centers. Exactly half of respondents (50%) expressed confidence, with 29.4% rating new graduates as "somewhat prepared" and 20.6% as "very prepared." The other half of the population has a cautious opinion, with 26.2% remaining neutral and 23.8% expressing concerns about graduates' readiness (21.4% "somewhat unprepared" and 2.4% "very unprepared"). Overall, the data suggests cautious optimism in the field, with most professionals seeing at least some potential in new graduates despite reservations about their complete readiness for this significant transformation.

### 6.6 Opinions Regarding Collaborative Spaces

The opinions regarding the collaborative spaces in libraries have been measured using Likert scale, the five point scale ranges from Strongly Disagree to Strongly Agree. Seven statements are measured and their distribution is given below in the Table 1.

**Table 1 Opinions Regarding Collaborative Spaces**

| Statements | SD | DA | N | A | SA |
|---|---|---|---|---|---|
| The traditional quiet library model is becoming obsolete | 4 (3.2%) | 19 (15.1%) | 34 (27%) | 39 (31%) | 30 (23.8%) |
| LIS professionals are adequately prepared for managing community library spaces | 4 (3.2%) | 26 (20.6%) | 31 (24.6%) | 63 (50%) | 2 (1.6%) |
| New library design should prioritize flexible, multi-use spaces | 4 (3.2%) | 6 (4.8%) | 8 (6.3%) | 54 (42.9%) | 54 (42.9%) |
| Balancing traditional and new library functions is a challenging practice | 2 (1.6%) | 20 (15.9%) | 27 (21.4%) | 65 (51.6%) | 12 (9.5%) |
| LIS education needs significant revision to address modern library functions | 1 (0.8%) | - | 5 (4%) | 64 (50.8%) | 56 (44.4%) |
| Current library architecture has provision for community places | 2 (1.6%) | 31 (24.6%) | 43 (34.1%) | 45 (35.7%) | 5 (4%) |

This table reveals strong support for modernizing libraries into collaborative community spaces, though with some notable concerns about professional preparedness. The most striking finding is overwhelming agreement (85.8%) that new library design should prioritize flexible, multi-use spaces, with nearly equal numbers strongly agreeing (42.9%) and agreeing (42.9%) with this approach. Similarly, there's robust consensus (95.2%) that LIS education requires significant revision to address modern library functions, indicating widespread recognition that current training may be inadequate. However, the data shows skepticism about current professional readiness, with only 51.6% feeling that LIS professionals are adequately prepared to manage community library spaces, while nearly a quarter (24.2%) disagree or strongly disagree. Respondents also express mixed views on whether current library architecture supports community functions, with 39.7% agreeing it does while 26.2% disagree. Interestingly, while 54.8% believe the traditional quiet library model is becoming obsolete, 18.3% still disagree, suggesting some resistance to completely abandoning traditional library concepts. The challenge of balancing old and new functions is widely acknowledged, with 61.1% agreeing this is difficult, reinforcing the complexity of this professional transition.

## 7. Hypotheses Testing

### 7.1 Hypothesis *H1*: Opinions regarding collaborative spaces are equal to average level

The hypothesis is tested using one one-sample t-test. Seven statements regarding library collaborative spaces along with their mean, standard deviation, t value of the response and p value is given below in the Table 2.

**Table 2: One-sample t-test**

| Statements | Mean | SD | t value | p-value |
|---|---|---|---|---|
| The traditional quiet library model is becoming obsolete | 3.57 | 1.11 | 5.8 | <0.001** |
| LIS professionals are adequately prepared for managing community library spaces | 3.26 | 0.91 | 3.22 | <0.001** |
| New library design should prioritize flexible, multi-use spaces | 4.17 | 0.97 | 13.56 | <0.001** |
| Balancing traditional and new library functions is a challenging practice | 3.52 | 0.93 | 6.25 | <0.001** |
| LIS education needs significant revision to address modern library functions | 4.38 | 0.64 | 24.1 | <0.001** |
| Current library architecture has provision for community places | 3.16 | 0.89 | 1.98 | <0.001** |
| How prepared do you feel new LIS graduates are to manage the transformation of libraries into communal centers | 3.44 | 1.11 | 4.48 | <0.05* |

**denotes significant level at 1% level          *denotes significant level at 5% level

From the table it is observed that the mean value is ranging from 3.16 to 4.38. The highest mean value (4.38) is for the statement “LIS education needs significant revision to address modern library functions” followed by mean 4.17 for the statement “New library design should prioritize flexible, multi-use spaces” and the least observed mean value (3.16) is regarding the current architecture of the library in accommodating community services.

Since p value is less than 0.01, the hypothesis is rejected at 1% level for all the statements of collaborative spaces except for the statement “How prepared do you feel new LIS graduates are to manage the transformation of libraries into communal centers” which is rejected at 5% level. Hence concluded that the opinion regarding all the statements of collaborative space is not equal to average level. Based on mean score, the opinion regarding the statements on collaborative space is above average level.

### 7.2 Hypothesis *H2*: There is no significant difference between the opinions of Gen Y and Gen Z.

This hypothesis is tested using t-test for difference of two mean or independent sample t-test. Seven statements regarding library collaborative spaces have been taken as criteria to compare and analyze the opinions of Gen Y and Gen Z. Mean and standard deviation of both Gen Y and Gen Z along with t value and p values are given below in the table Table 3.

**Table 3: Independent sample t-test**

| Statements | Gen Y (Before 1996) | | Gen Z (After 1996) | | t value | p-value |
|---|---|---|---|---|---|---|
| | Mean | SD | Mean | SD | | |
| The traditional quiet library model is becoming obsolete | 3.34 | 1.15 | 3.76 | 1.04 | 2.16 | 0.033* |
| LIS professionals are adequately prepared for managing community library spaces | 3.5 | 0.8 | 3.06 | 0.96 | 2.77 | 0.006** |
| New library design should prioritize flexible, multi-use spaces | 4.31 | 0.79 | 4.06 | 1.09 | 1.45 | 0.149 |
| Balancing traditional and new library functions is a challenging practice | 3.59 | 0.99 | 3.46 | 0.87 | 0.79 | 0.434 |
| LIS education needs significant revision to address modern library functions | 4.31 | 0.65 | 4.44 | 0.63 | 1.14 | 0.257 |
| Current library architecture has provision for community places | 3.21 | 0.87 | 3.12 | 0.92 | 0.55 | 0.58 |
| How prepared do you feel new LIS graduates are to manage the transformation of libraries into community centres | 3.79 | 1.06 | 3.15 | 1.08 | 3.38 | 0.001** |

**denotes significant level at 1% level *denotes significant level at 5% level

Since p value is less 0.01, the hypothesis is rejected at 1% level with regard to statements of collaborative space (LIS professionals are adequately prepared for managing community library spaces & How prepared do you feel new LIS graduates are to manage the transformation of libraries into community centres). Hence there is significant difference between Gen Y and Gen Z with regard to these statements of collaborative space. Based on the mean score, Gen Y has better opinion on these statements than Gen Z. It can be assumed that the better opinion of Gen Y is because they have had more time to develop practical skills to manage community integration programs of the library. The years of experience made them witness the gradual transformation of libraries. The technological integration into libraries has been a smooth transition and it made the Gen Y professionals confident about the management of the transformation of the libraries.

Since p value is less than 0.05, the hypothesis rejected at 5% level with regard to statement "The traditional quiet library model is becoming obsolete". Hence there is significant difference between Gen Y and Gen Z with regard to this statement of collaborative space. Based on mean score, Gen Z has better opinion on this statement than Gen Y. It can be interpreted that the Gen Z opinion regarding the obsolesce is because they grew up in a technologically immersed world and traditional library model does not full fill the technology and social needs of Gen Z.

There is no significant difference between Gen Y and Gen Z, since p value is greater than 0.05. Hence hypothesis is accepted at 5% level with regard to statements "New library design should prioritize flexible, multi-use spaces", "Balancing traditional and new library functions is a challenging practice', "LIS education needs significant revision to address modern library functions" and "Current library architecture has provision for community places". Both generations Y and Z understand the importance of evolution of libraries as well as difficulties came along with the transformation.

## 8 Conclusion

The findings of the study indicate a shift in the perspective of the library professionals. Change is importance in the case of service oriented institutions, changing time and technology demand improvised services. The traditional restricted service model is changing to dynamic community centered models, social gatherings and community building programmes will shape the libraries into backbone of society and democracy.

Opinions of previous 10 batches of MLISc graduates encompasses gen Y and Gen Z were taken for analyzing the importance of collaborative space and community engagement programmes in the library. The result of the study revealed that all the statements about collaborative spaces are above average level that is there is a positive attitude towards incorporating community spaces into current library services. Comparison part of the study resulted that in the case of preparedness of library professionals in welcoming community services and obsolesces of LIS curriculum there is a significant difference between Gen Y and Gen Z.

Dr. S R Ranganathan's fifth law of library science that is "Library is a growing organism" pointed out the need for the evolution and updating of library services from time to time. Being a service oriented institution incorporating new services is important to provide efficient services. Community spaces and collaborative services are the demand of the present era and as the positive nature of the result prove the theory. More over library should be place where people escape from digital environment and build socio political culture.

## References

Garoufali, A., & Garoufallou, E. (2022). Transforming libraries into learning collaborative hubs: The current state of physical spaces and the perceptions of Greek librarians concerning implementation of the "Learning Commons" model. Global knowledge, memory and communication, 73(6/7), 828-852.

Ramya, A. V., & Majeed, A. (2025). Public Libraries as Community Engagement Hub: An Analytical Study. South India Journal of Social Sciences, 23(2), 26-30.

Abshana, J., Joseph, M., & Vijesh, P. (2023). A public library as a gateway for social work community practice: An information-inpractice from Kerala, India. Journal of the Australian Library and Information Association, 72, 492–499. https://doi.org/10.1080/24750 158.2023.2271694

Abu, R., Grace, M., & Carroll, M. (2011). The role of the rural public library in community development and empowerment. International Journal of the Book, 8, 63–74. https://doi.org/10.18848/14479516/ CGP/v08i02/36863

Barath, P. S., & Sudhier, K. G. (2024). Rural public library's outreach services in bridging the digital divide in Thiruvananthapuram district: A study on librarian's perspectives. In AI-assisted library reconstruction (pp. 256–266). IGI Global.

Hapel, R. (2020). Building a contemporary public library through community engagement: A personal journey. Public Library Quarterly, 39(5),391–409. https://doi.org/10.1080/01616846.2020.1772621

Koteswara Rao, M. (2019). Public libraries in India: Problems and prospects. Pearl: A Journal of Library and Information Science, 13.https://doi.org/10.5958/09756922.2019.00034.2

Krejcie, R. V., & Morgan, D. W. (1970). Determining sample size for research activities. Educational and Psychological Measurement, 30(3), 607-610.

Kranich, N. (2012) "Libraries and Civic Engagement", Rutgers University Community Repository. http://dx.doi.org/doi:10.7282/ T3VX0DWS

Kunhambu, K., & Mudhol, M. V. (2012). Role of rural libraries in rural areas. SRELS Journal of Information Management, 49(6), 631–647. https://doi.org/10.17821/srels/2012/v49i6/43920

Larsen, H. (2024). Managing Norwegian public libraries as civil public spheres: Recent controversies. Journal of Documentation, 80(1), 116– 130.

Lee, P. C. (2023). From Community Engagement to Community Resilience: The Evolving Role of Public Libraries. Public Library Quarterly, 43(3), 339–366. https://doi.org/10.1080/01616846.2023.2 251860

Monroe-Gulick, A., Weaver, M. D., & Morris, S. E. (2024). Women and men in library and information science scholarship: Authorship trends from 2003 to 2021. *The Journal of Academic Librarianship*, *50*(5), 102939. https://doi.org/10.1016/j.acalib.2024.102939

Reid, H., & Howard, V. (2016). Connecting with community: The importance of community engagement in rural public library systems. Public Library Quarterly, 35(3), 188–202. https://doi.org/10.1080/016 16846.2016.1210443

Van Melik, R., & Merry, M. S. (2021). Retooling the public library as social infrastructure: A Dutch illustration. Social & Cultural Geography, 24(5), 758–777. https://doi.org/10.1080/14649365.2021. 1965195

V.Calista., & Suharjanto, G. (2024). Enhancing sustainable public library: Spatial relationship implementation to foster optimal social interaction. https://doi.org/10.1088/17551315/1324/1/012011